# Enterprise Resource Planning - 'Real blessing' or 'a Blessing in Disguise': An Exploration of the Contextual Factors in Public Sector

[1]Shafqat Ali Shad (shafqat@mail.ustc.edu.cn),

[2]Enhong Chen (cheneh@ustc.edu.cn)
Department of Computer Science and Technology
University of Science and Technology of China
Huangshan Road, Hefei, 230027 Anhui, China

[3]Faisal Malik Faisal Azeem (faisal.azeam@comsats.edu.pk)
COMSATS institute of information technology, Pakistan
CIIT, Quaid Avenue Wah Cantt, Pakistan

# Enterprise Resource Planning - 'Real blessing' or 'a Blessing in Disguise': An Exploration of the Contextual Factors in Public Sector

**Abstract**

Information systems have always been in a prime focus in organizations in both local (Pakistani) and global environment. Now the race of being the best through Information Systems has created its importance in public sector organizations to meet the global challenges. Public sector organizations have been facing problems in different segments of technology adoption especially in ERP projects. ERP adoption/implementation projects in public sector organizations still encounter major setbacks in terms of partly/completely success/failure. Cultural and other social barriers have been resistant in technology adoption in Pakistan. Now in the case of big ERP adoptions the contextual factors must be identified and addressed. The paper investigates the reasons of success or failure by addressing nature of complexities regarding different contextual factors. The study includes a sample of Pakistan's four public sector organizations. The sample of this four organizations includes two organizations (Type-A) i.e. Oil & Gas Development Company Limited (OGDCL) and National Database Registration Authority (NADRA) where ERP has been successfully implemented and other two (Type-B) i.e. Pakistan Telecommunication Corporation Limited (PTCL), Higher Education Commission (HEC) where ERP implementation is in progress. The findings address the contextual factors i.e. cultural, environmental & political changes which have a variable impact on ERP systems adoption/implementation in addition to Business Process Re-engineering (BPR). Paper also briefly includes analysis of gaps between pre & post ERP implementation scenarios.

1. **Introduction**

Government of Pakistan (GoP) has started making its owned organizations highly competitive in order to meet country's internal and global information communication requirement reliably and in a secure manner. Since the last decade GoP starting putting her efforts towards technology in almost all the industrial and community life sectors both in terms of infrastructure and Skill development. Now GoP is moving towards successful ERP adoption in its industrial sectors like Banking, Telecommunication, Education, Logistics and Energy etc. State Bank of Pakistan (SBP) started drawing ERP benefits though by encountering many problems since its ERP project initiation in July, 2002 but rest of the sectors are still into the process of ERP adoption/implementation. GoP is putting billions of dollars onto implementation of ERP solutions by having a standard world class ERPs with internationally recognized teams of consultants. The Study limits itself on four public sector organizations (Type-A & Type-B) i.e. Pakistan Telecommunication Corporation Limited (PTCL), Higher Education Commission (HEC), Oil & Gas Development Company Limited (OGDCL) and National Database Registration Authority (NADRA). The paper highlights the reasons of ERP failure by focusing upon the gaps that arise other than the areas i.e. cost and time but the contextual factors that are generally ignored or least focused but may be the reasons of project failure. The gap which may become cause of the effects on the successful ERP adoption and implementation is explored in the paper. In Pakistan's scenario because of not having sufficient knowledge in this domain organizations may suffer in identification and treatment of the contextual factors in ERP projects. The reason of missing these factors may be due to the following predefined international standard practices as accomplished in other developed or developing countries that may vary from region to region and country to country. This paper may work as a facilitator for the management of public sector organizations of Pakistan who are involved in ERP implementation and for those who are planning to move towards ERP adoption.

The paper follows a particular sequence, as in the coming section paper briefly describes the relevant research in developing countries in public sector organizations. Section III describes the research design by investigating ERP implementations in the three above stated sample organizations of Pakistan. The fourth section portrays the research environment. Section-V interprets the findings through the data analysis and at the end section VI will be concluding the findings through analysis of a CSF model. The research will pave the way for the managers to decide proactively about what are the most critical contextual factors and their handling that may convert their ERP projects into a "true blessing" or a "blessing in disguise".

**2. Enterprise Resource Planning (ERP) systems**

Since the evolution of ERPs in 1990s and then of extended ERPs in 2000s (Mohammad et al, 2002) it is considered as an instant replacement of the legacy systems and operating processes (Mullin R. 1999). Any ERP package implementation requires extensive business process reengineering and alignment (BPR & A) (Holland et al,1999) so that the gaps of perceived and actual happenings may be reduced. Different well known ERP systems like SAP, Oracle, People Soft, JD Edwards and BAAN have penetrated in the market because of their extensive features and potential market needs. But ERP systems according to Sally Wright, Arnold M. Wright, (2002), always bear high risk due to its cross functional inter-relationship with business processes. This relationship if not clearly defined and communicated among all the stake holders can cause many contextual complexities that may lead the project towards failure. ERP system implementation complexity, drastic cultural human and organizational changes and at times the high customization cost drive the customer to go over the implementation plan carefully (L. P. Willcocks and R. Sykes, 2000).

As in the public sector of Pakistan, Out of the sample of four, when two organizations of Type-A have been explored initially. HEC (Higher Education Commission) of Pakistan in 2006 contract awarded to the Siemens Pakistan Engineering Company for mySAP based modules implementations i.e. FI/CO (Finance and Controlling), HR ( Human Resource Management) MM ( Material Management) and HEC Project System (Anwar & Amjad, 2007). HEC keenly showed interest in skill development through training of SAP modules to its employees for making this ERP project successful. Pakistan Telecommunication Ltd the largest telecommunication company in Pakistan being managerially controlled by Etisalat International Pakistan LLC, since April 12, 2006 by acquiring its 26% shares (www.etisalat.ae retrieved as on 12-12-2009). Implementation of SAP-based ERP in the area of Financial Management and Supply Chain Management, Human Capital Management, Network Lifecycle Management, Planning and Business Process Optimization with the cost of 1.6 billion is on the go by Siemens Pakistan. Project will prolong for 20 months with momentous deliverables every 6 months (Zubair Qureshi, 2007). Type-B organizations of the selected sample are with ERP implemented into them i.e. OGDCL & NADRA. Oil and Gas Development Company Limited (OGDCL) working on Oracle Financials being one of the early ERP adopters in public sector of Pakistan, they have raised their annual sale growth up to Rs.100.26 billion (2006-2007). NADRA is considered as one of the critical implementations because of its organizational and the nature of work to be accomplished. A large scale implementation of Oracle ERP Suite has been made in NADRA in addition to a leading human-centric BPM platform Ultimo. Addressing the layers of culture (W. Skok and H. Döringer 2001) in any organization, disruptive organizational changes evidently occur when ERP implementation moves on (Soh et al, 2002). Human behavioral, cultural and social attributes need to be satisfied in contrast with the type and level of changes through ERP adoption in any organization. The suggested reasons of ERP failure are not just technical issues but concerned behavioral factors as well (W. Skok and H. Döringer, 2001). According to Chatfield there is a noteworthy impact of Organizational culture and structure on ERP implementation (C. Chatfield, 2000). Technology implementations need an assessment of human management and organizational risk to mitigate the technical risk for the system success (D. L. Olson, 2001). Contextual factors that contribute in the failure of ERP implementation highlighted as lack of management commitment, end user involvement in the project, poorly defined communication management structure, political intrigue and hidden agendas (Stein et al, 1999). Roles and responsibilities of the workforce surely change in ERP adoption process which should be pre-defined and set up (Rodney et al, 2005). Zhenyu Huang & Prashant Palvia, 2001 sate that "A company with a strong culture would have better understanding of application functionality, data management, and more accepting of ERP systems" (Zhenyu Huang & Prashant Palvia, 2001) which reflect towards the re-organization of culture, political and environmental norms in a proactive manner. The success of an ERP project is settled upon the acceptance of the changes by organization /employees i.e. organization's cultural changes or Business Process Redesign (BPR) instead of the reasons of system bad performance and cost etc (Kenneth et al, 2002) In Pakistan as interviewed from around 30 respondents (employees) general perception prevailed in the public sector employees that is IT implementations are though being going on but these are dominated and affected by corporate culture, beaurocratic behavior of the top level management and political structure and observed as the strongest barriers in its success. So as stated by

Hsin Hsin Chang, 2006 that corporate culture dominates the IT functions to play its due role strategically (Hsin Hsin Chang, 2006).

Critical success factors have been analyzed by a lot of researchers in the past as Slevin and Pintor (1987) talks about tactical and strategic project management capabilities, Gibson (Holland et al, 1999) states about the top management support and Holland at all (Holland et al,1999) narrate deep emphasis on the need of project phases keeping in view the individual actions in an ERP project implementation. Leadership also plays a critical role in the success of an ERP project (Sarker and Lee S. Sarker and A. S. Lee, 2000). Critical Success factors have been pragmatically analyzed by Christopher P. Holland and Ben Light presented in their CSF model which has also been used and applied in this paper. For the purpose of going for the Implementation of ERP, strategic and tactical approaches are used which are strategically important to be considered. The model stated below covers the same especially ERP projects need to include these factors to be analyzed.

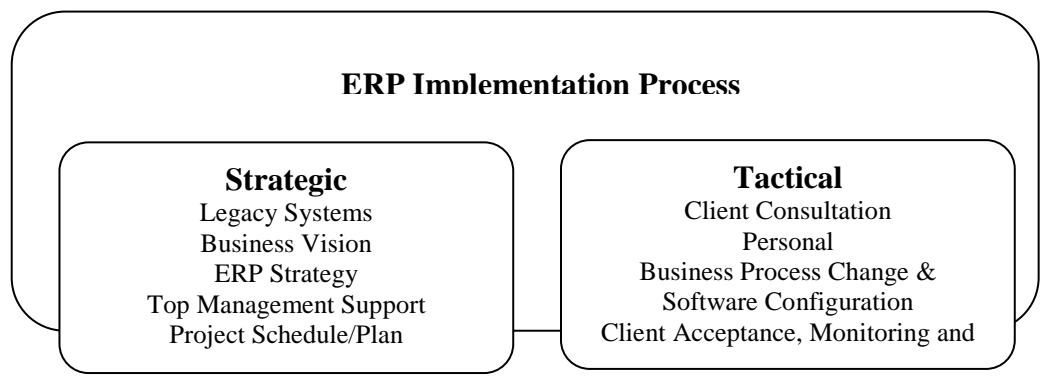

**Figure 1: A Critical Success Factors Model of Implementation from Holland et al. (1999) and later on by David Allen, Thomas Kern & Mark Havenhand (2000)**

The above stated CSF Model later on experimented through case study method successfully. Now this paper pragmatically analyzes the contextual factors i.e. cultural, environmental and political in addition to other tactical and strategic ones in public sector organizations of Pakistan where maximum number of ERP solutions is being and have been adopted as off-the-shelf from different international ERP vendors.

## 3. Research Methodology

As the prime objective of the research study is to explore in the impact of the real contextual factors that affect public sectors' ERP implementations in Pakistan both in a positive and negatively. The research is exploratory in nature with statistical data analyses gathered from four public sector organizations of Pakistan unlike the historical researches which have mostly been made by case study methods. There is no prior research in this context in especially in this region so the generalizability of this study has been taken into consideration in ab-innitio by investigating the ERP implementations in public sector organization and past researches have also been taken into account. The study focuses upon the contextual factors stated above, in the light of CSF model of Holland et al.

Sample of four public sector organizations has been selected and the study explored the ERP implementations in public sector organizations of Pakistan taking into account the contextual lens of CSF model presented earlier by Slevin & Pinto (1987) and Holland et al (Holland et al,1999). When started exploring ERP implementations in Pakistan, the researcher found quite stubborn and non-cooperative environment for research purposes then the study made through a sample based data collection from the employees of four public sector organizations and reached at the findings at the end from the collected data through questionnaires and informal interviews. The three independent variables (contextual factors) have been analyzed for their affects on ERP implementations in public sector of Pakistan. For this study officials of the organizations have been contacted, most of them were reluctant in providing information because of prevailed potential threats in their minds and off course due to their organizational structure/culture. Finally four organizations were selected for exploration of the contextual factors for ERP implementations i.e. HEC (Higher Education Commission) & PTCL (Pakistan Telecommunication Company) (Type-A) for pre implementation analysis and OGDCL (Oil and Gas Development Corporation) & NADRA (National Database Registration Authority) (Type-B) for post implementation analysis. In addition to the contextual factors stated above this study also included briefly the strategic and tactical areas in order to be more realistic and authentic in research findings. The data collection undertaken for this research is primarily through questionnaires and informal interviews and through the literature known/available about these organizations.

**4. The Research Environment**

In Pakistan, Public sector organizations have their own typical hierarchical structure. They have historically been facing a lot of contextual barriers in acquiring technology. Though they put their emphasis on efficiency initially but as the directions of the global market towards technology changed their pace, they also had to move towards big IT implementations. As a result ERP (Oracle Financials) in OGDCL and SAP in central bank i.e. State Bank of Pakistan successfully implemented by the government though they faced a lot of tactical and strategic barriers. Later on ERP implementations have been encouraged into other public sector organizations by the Government. In the last decade private sector giants due to the prevailed and forth coming market opportunities in Pakistan also started shifting themselves from the old complex IS structure to ERP based environment especially in Telecom, Banking and Energy sectors. Being a new market for ERP vendors like Oracle Financials and SAP put a lot of emphasis on it and are successfully moving towards their business objectives. Pakistan Software Export Board is also putting its endeavors towards the development of ERP solutions both open source and licensed for the SMEs and large organizations. The research focuses upon the Off-The-Shelf ERP solutions in large public sector organizations because of the need to analyze the real contextual barriers that cause delays and failure of ERP projects. As analyzed the complexity of the offered ERP packages in Pakistan, these packages do have a lot of complexities but there are more complex contextual barriers in Pakistan that lead them more towards failure. Need of information for the government in different its sectors has been increasing and this local and global pressure getting more and more enhanced and pushing government to put more emphasis on big technological and structural investment in the form of ERP implementation especially.

**Hypothesis postulated**

$H_1$: ERP implementation is drastically affected by contextual factors (i.e. cultural, environmental and political) with their effects on each other in public sector organizations in Pakistan

**5. Analysis and Discussion of the Critical Success Factors**

The research mainly focuses upon the identified factors of that has earlier been exposed by Holland and Light (Holland et al,1999) and later on by David Allen, Thomas Kern & Mark Havenhand [21] which states two types of tactical and strategic factors which need to be addressed when assessing ERP implementation. When these factors have been dugout in the current environment for the purpose of analysis were found quite critical as the role of legacy systems and old process design must be analyzed due to the practicing activities that ultimately reflect towards the new technology i.e. ERP adoption and new strategy development. Holland and Light (Holland et al,1999) interpret business and IT systems as a pool of existing business processes, organizational culture, and organization structure which are the determinants of the amount of IT and organizational change required to implement ERP successfully. The model though reflects a true picture but when researched the contextual factors, drills down the concept of socially constructed organizational conditions that may become the main reasons of success or failure of ERP implementation in public sector organization (David Allen, Thomas Kern & Mark Havenhand, 2002). There may be the difference in explanations but as a whole intensity of these contextual factors has been found quite critical which if not addressed can lead the project towards failure. This paper focus is though primarily on contextual factors but all the elements described in the CSFs model stated above were brought into analysis as well but these factors have been analyzed in connection with the contextual factors. Fig-II uncovers the contextual factors being part of the CSFs model that shows the behavior of these factors as a lens for successful ERP implementation.

**Critical Success Factor model of ERP**

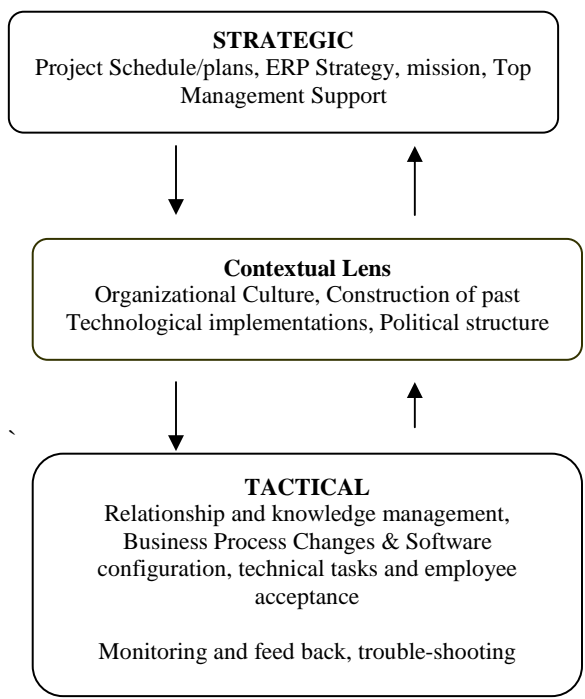

_______________________________________

**Implementations adapted from Slevin & Pinto (1987)
[16], Holland et al. (Holland et al, (1999) and David Allen et al (1999)**

Prior to the current research same model has been applied on public sector higher education institutions (HEIs) by David Allen et al using case method which is infect the extension of some of the areas of Holland and Light's (Holland et al,1999) research efforts. This research paper focuses on the same areas but with a different approach and additional contextual factors.

Data collected from four Pakistani public sector organizations through face to face informal interviews from their employees and through personally delivered questionnaires. Informal interviews contained open ended questions regarding their view point about ERP, culture, environment, their views about organizational internal & external political structure and their views about organization's top level management without following a planned sequence of questions. The survey questionnaire was same for all four (sample) organizations to have a view of their pre and post implementation scenarios regarding contextual factors stated above. It consisted of three types of formats. Professional & personal information (general format), scaled response questions and dichotomous questions. At the time of data gathering the researcher face a very low degree of cooperation from the employees due to various contextual issues. It has been intensively observed that the employee there were quite reluctant in providing required information especially in written questionnaires though they responded in a comparatively better manner in verbal informal interviews because of considering this information leakage as a threat to their jobs. Initially this questionnaire was tested in all four public sector organizations in a low profile in order to get early feed back and response. Then it was redesigned and distributed in all four selected organizations. Most of the employees found unaware of ERP, its actual purpose and its features even though they are using it, especially in NADRA with Oracle financials implemented. They quoted as they are provided an ERP number through which they perform limited tasks required by the organization. Those who provided their viewpoints, they did not provide information from which they may be traced. OGDCL which is Pakistan's one of the early ERP (Oracle Financials) adopters found comparatively good among its employees regarding its ERP. In the same manner when the study went through the data collection phase at HEC it was quite difficult for the researcher to find out the people there who have been part of ERP/SAP project or those who know about ERP implementation in progress in their organization for the purpose of interviews and questionnaire response. Employees in PTCL were quite a bit reluctant in providing required information especially in written form because of the political and environmental issues within and outside the organization (As mentioned by many of the employees there) amongst employees and the management. 100 questionnaires had been distributed in each of the public sector organizations listed above. Due to the political, cultural and environmental factors discussed above, 30 questionnaires had been received back from the respondents from each organization except PTCL from where 33 returned. In order to analyze the data, tools used were SPSS and MS excel.

Data collected through the questionnaires and informal interviews after analysis, reflected in three formats (As the structure of the questionnaire and interview was designed) lead the study towards the same direction. The Cronbach's alpha of the both types of organizations, one where ERP implementation is in progress and the other with ERP implemented in itself derived from SPSS was [0.825] & [0.873] respectively. 82.5% & 87.3% ensured the reliability of the study. [Details attached in Table-1 in Annexure] After the analysis of the dichotomous questions part, reasons of ERP project delays primarily are lack of planning, lack of management skills, slow decision making and lack of technical skills which are 88.33333%, 70%, 66.66667% and 68.33333% respectively [table-3, annexure]. Behavioral and Authority related concerns were primarily highlighted. Lack of skills as a major threat and existing culture as one of the prominent barriers identified.

**Table-2**

| Contextual Factors | OGDCL | PTCL | NADRA | HEC |
| --- | --- | --- | --- | --- |
| **Culture** | (23.727) | (42.236) | (41.958) | (26.093) |
| p-value | .000000 | .000000 | .000000 | .000000 |
| **Environment** | (33.463) | (40.652) | (41.865) | (29.976) |
| p-value | .000000 | .000000 | .000000 | .000000 |
| **Political** | (47.132) | (34.077) | (36.666) | (24.319) |
| p-value | .000000 | .000000 | .000000 | .000000 |

[*The values shown in the parentheses in the table-2 are the t-values]

All four organizations when analyzed that t-value is high, and the p-value is .000000 of all the factors for each organization which reflects as a strong evidence against the null hypothesis(Ho).As on the basis of t-test and significance level drawn, we can safely reject the null hypothesis and will accept the alternate one. It means ERP implementation is drastically affected by contextual factors (i.e. cultural, environmental and political) with their effects on each other in public sector organizations in Pakistan. Details have been provided in table 1a in annexure.

**Organizational Culture & environment**

Culture as a whole affects on the productivity of the organizations by affecting individuals and drastically affects the thinking of the employees in any organization (Bandura, A. (1977) (1982). Culture as "a pattern of basic assumptions from groups of individuals derived from different perceptions" (Schein, E. H. 1985) contributed both positively and negatively on the happenings in any organization. Early researches state that cultural changes due to any of the organization change like BPR, ERP or TQM includes employees' behavior and their efficiency can cause good or bad in any organization (Al-Khalifa et al, (2000), Hoffman et al, (2000). This paper focuses upon the cultural factors that are affected by the perceived change and ultimately affect individual perceptions regarding ERP implementation in public sector organizations in Pakistan. Communication among the stakeholders regarding ERP implementation effects form their beliefs and perceptions if done in a right and planned manner (Bates et al, 1995). ERP implementations in relation with cultural factor have already been explored and analyzed (Taylor, et al, 1998) so, is being preceded in this paper by focusing only on public sector organizations. As analyzed in Pakistan's public sector organizations it do & did affect ERP implementation projects in the form of delays, cost and time overruns etc. No doubt since the past decade Government of Pakistan has changed her approach towards technology adoption by following a global and private sector patterns but faced a lot of hindrances in doing so not just only because of required skill set and complexity and cost of the software/ERP but due to the contextual factors especially cultural ones that had not been positively addressed.

At the time of data collection through questionnaires and informal interviews in two public sector organizations where ERP implementation is in progress i.e. HEC and PTCL, many of the employees were threatened by the perceived drastic cultural change in the organization. This unrest when analyzed was found due to neglecting the employees concerns and by considering culture as a non-affecting factor which can easily be re-enforced in any organization. The culture in Pakistan public sector organizations as observed and interviewed is though inflexible but need to be addressed rightly when BPR for ERP and ERP Implementation itself or any other change major is required. Most of the respondents (employees) during the interviews have been found unaware of the ERP even by being end users of it. According to some of the respondents in NADRA, they don't know what ERP is, they are just provided an ERP number as their ID, they just view and use their screen as trained up to a limited scope/scale.

i. In the data collection phase the below stated issues regarding environmental and cultural issues had been addressed in both pre & post ERP implementation scenarios in public sector organizations of Pakistan i.e.

ii. Organization culture i.e. its norms and rules

iii. Attitude of employees towards work

iv. Employee selection and involvement criteria for ERP project team

v. Change in employee behavior/attitude

vi. ERP as a social threat

vii. BPR for ERP implementation

viii. Social needs fulfillment

ix. Employees reluctance due to dual responsibilities

x. Change in management structure (hierarchical etc)

xi. ERP Implementation as an opportunity

xii. Financial benefits before & after ERP implementation

xiii. General environmental change acceptability through ERP

xiv. Organizational structure flexibility for ERP implemented

xv. Change in working standards through ERP

xvi. Monitoring threats after ERP implementation

**Political Structure**

Bureaucracy, during the course of study, when drilled down in German sociology (Weber and Max , 1962) found as one of the most famous approaches of organizational governance. As observed from the existing literature in today's knowledge base economy a typical bureaucracy has been changing its shape from a rigid control base over employees and processes to knowledge based flexible technology, business and management base. This culture/structure has been used by many large and complex organizations round the globe by encouraging the already tested and practiced moves which no doubt behaves as a barrier in BPR for ERP implementation. Senge, (1990) states about the capabilities of continual change of the organizations which work in an unstable or highly volatile environment. When talk about public sector organizations of Pakistan their top level management is highly influenced or supposed to be answerable to the state representatives/rulers so they are controlled politically to great extent. HEC and PTCL employees when interviewed, they reported a heavy involvement of rulers (As Pakistan is facing a political unrest for last many years) which caused so many problems and ambiguities in ERP implementation projects. Even trying to be flexible the public sector organizations followed the private sector technology adoption and usage patron but still the political structure has developed strong roots into it due to which enforcement of any change in any of these organizations takes the productivity down and as a result the unattended problems and misconception in the minds of the employees/stakeholders lead that particular change towards failures. The ultimate decisions though are made in the high level groups of individuals but are not properly communicated down the line amongst the employees who carry forward these activities with a lot of risks.

Data gathered in this regard focusing on the below stated issues in two different ERP implementation (complete & in progress) scenarios in public sector organizations of Pakistan i.e.

i. Bureaucratic management style

ii. End user involvement in different phases of the project

iii. Training (formal & informal)

iv. Misconceptions / fears like rightsizing or downsizing, loss of authorities etc addressed by the Top Level Management

v. Political influences of regulatory bodies and the government on the public sector organizations

vi. Conflicts resolution and management

vii. Potential risk(s) awareness associated with ERP implementation

viii. Communication planning and management

ix. Criteria for selection of ERP vendor and consultant for implementation

**Social construction of technological legacy**

Legacy systems play a vital role in organizational strengths but these technologies may often cause failure of the organizations because of the obsolescence issue(s) and volatile market conditions. At times organizations are distressed by legacy technology failures (D. Knights and F. Murray, 1994) so as the happening in Pakistan. From the collected data analysis averages derived from the particular question of BPR, an overall lickert scale average of 3.269841 and 3.051724 were drawn out of the employees' response for both (Pre & post implementation) types of organizations respectively which theoretically reflects that majority of the employees is quite uncertain or to an extent in favor of change in exiting/traditional organizational processes. In Pakistan's public sector organizations since the last decade technology penetrated in a very creepy manner because of unawareness, poor technology infrastructure, potential threats amongst stake holders, political and economic instability etc Even the issue of process obsolescence have been ignored for many years due to being uncontested in the market place within the country. Now it's a matter of survival in a highly competitive market within and outside the country they are bound to be best in class to serve the public. When interviewed from randomly selected respondents from different public sector organizations it was observed that they were socially and skill wise threatened by the technology revolution through ERP in their organizations although they want process change that's why they are not ready to switch from the legacy systems to the advanced ones.

**Relationship and knowledge management**

Early research (S. Chang, 2000) highlights the criticality of ERP in public sector organizations because of its highly complex nature both in terms of technical and managerial contexts. Conesus between all the stake holders is quite critical for successful ERP implementation in public sector organizations. Knowledge gaps are fulfilled by mutual understanding, required skill set and implementation experiences the different parties involved in ERP implementation. From the data collected through the questionnaires and informal interviews, the level of awareness and clarity about ERP was not up to the mark. Employees in one of the selected organizations were not clear about what could be the maximum output of what their job is/will be before and after its implementation. The research overall derived a conclusive overall applied scale average of 3.492063 in the organizations where ERP implementation is in progress and 3.533333 where ERP has already been implemented. The response evidently shows that there is high need to involve employees in order to avoid delays and cost overruns in ERP implementation projects.

**Communication as a political process**

Study in the phase of data analysis explored an average communication structure in ERP projects in public sector organizations in Pakistan. Though there is a strong will and interest of the top level management in ERP implementation in all analyzed organizations of Pakistan but their approach traditionally was more towards enforcement instead of positively convincing. Now the top management is trying to address the communication needs within the organization and the ERP project but still not reached up to the required level of satisfaction which as a result comes up as one of the contextual barriers. Many of the respondents showed their deep concerns about "limited sharing approach" in their organizations though they are informed directly or indirectly to an extent. Those who are involved generally don't share any or all the information with their colleagues considering it a threat for their job from the top level management and from their peers as well. Slow decision making process has been complained by the respondents is just because of the complex hierarchical decision structure. When some or all the concerns of the stake holders, especially internal stakeholders are not addressed, they react in different ways is the forms of rumors of threats and project failure most of the times even before ERP implementation. High ups generally ignore the real importance of communication across the organization in public sector and face very critical problems which finally lead them towards project failure.

## 6. Conclusions

This research paper is a continuity of the previous efforts made in the area of ERP implementation pros and cons. In Pakistan no such effort on ERP in Public sector was found earlier especially when GoP is putting a lot of its resources towards technology implementations in its owned organizations across different industries. As the paper focus is on the contextual factors i.e. culture, environmental and political, in addition to the brief view of BPR, communication process, legacy systems and relationship with knowledge management. It comes with the point of view that in addition to the core focused areas of ERP implementation, public sector organizations bear more risk as compared to the privately owned bodies because of their strongly rooted cultural, environmental and political structure. These contextual as proved from the data analysis gathered from public sector organizations of Pakistan are one of the major causes of project delays, cost and time over run or complete project collapse. Where ERP implementation is partly or completely successful even these organizations are still facing problems of distrust from their employees and unable to create loyalty with them. The framework provided (Holland et al,1999) has paved the way for its application onto the neglected areas which most of the times place barriers in ERP success. Pakistan is passing through critical political and economical crises which in tern are causing both short and long term loss. ERP projects are though tried to be run on private sector pattern but with different methodology of handling its internal stake holders. This exploratory research nullify Ho by proving that the ERP implementation is drastically affected by contextual factors (i.e. cultural, environmental and political) with their effects on each other in public sector organizations in Pakistan. When gap between the two types of organizations analyzed, intensively found lack of proper planning and stake holders' involvement due to which Type-A organizations have suffered with the issues of time delays and cost overrun. Type-B organizations with ERP implementation in progress are doing although better but repeating the same pattern of ERP adoption which reasoned critical problems for the early ERP adopters. Beaurocratic management style placed a lot of hurdles in the success of successful ERP implementation stated by the employees of Type-B organizations and the same practice has been observed in Type-B with a slightly less proportion. Threats of social disintegration through ERP are comparatively more in Type-B organization. Majority of the employees of Type-A organizations viewed that there is no drastic change positive change after ERP implementation where as employees of Type-B organizations are quite optimistic in this regard. If the addressed contextual factors are properly served, the adoption/implementation of ERP will be a collective effort of all the stake holders which ultimately result in its success.